\documentclass[journal,twoside,web]{ieeecolor}
\usepackage{generic}
\usepackage{cite}
\usepackage{amsmath,amssymb,amsfonts}
\usepackage{algorithmic}
\usepackage{graphicx}
\usepackage{textcomp}
\usepackage{tabularx}
\usepackage{multirow}
\usepackage{subcaption}
\usepackage[ruled,linesnumbered]{algorithm2e}
\newcommand\setrow[1]{\gdef\rowmac{#1}#1\ignorespaces}
\newcommand\clearrow{\global\let\rowmac\relax}
\def\BibTeX{{\rm B\kern-.05em{\sc i\kern-.025em b}\kern-.08em
    T\kern-.1667em\lower.7ex\hbox{E}\kern-.125emX}}
\clearrow
\begin{document}
\title{Generative Adversarial U-Net for Domain-free Medical Image Augmentation}
\author{Xiaocong Chen, Yun Li, Lina Yao, \IEEEmembership{Member, IEEE}, Ehsan Adeli, \IEEEmembership{Member, IEEE}, \\and Yu Zhang, \IEEEmembership{Senior Member, IEEE}
\thanks{Xiaocong Chen, Yun Li and Lina Yao are with the School of Computer Science and Engineering, Faculty of Engineering, University of New South Wales, Sydney, NSW, 2052, Australia. (e-mail: \{xiaocong.chen,yun.li5,lina.yao\}@unsw.edu.au)}
\thanks{Ehsan Adeli is with the Department of Psychiatry and Behavioral Sciences, Stanford University, Palo Alto, CA 94305 USA (email: eadeli@stanford.edu).}
\thanks{Yu Zhang is with the Department of Bioengineering, Lehigh University, Bethlehem, PA 18015, USA.(e-mail: yuzi20@lehigh.edu).}
}

\maketitle

\begin{abstract}
The shortage of annotated medical images is one of the biggest challenges in the field of medical image computing. Without a sufficient number of training samples, deep learning based models are very likely to suffer from over-fitting problem. The common solution is image manipulation such as image rotation, cropping, or resizing. Those methods can help relieve the over-fitting problem as more training samples are introduced. However, they do not really introduce new images with additional information and may lead to data leakage as the test set may contain similar samples which appear in the training set. To address this challenge, we propose to generate diverse images with generative adversarial network. In this paper, we develop a novel generative method named generative adversarial U-Net , which utilizes both generative adversarial network and U-Net. Different from existing approaches, our newly designed model is domain-free and generalizable to various medical images. Extensive experiments are conducted over eight diverse datasets including computed tomography (CT) scan, pathology, X-ray, etc. The visualization and quantitative results demonstrate the efficacy and good generalization of the proposed method on generating a wide array of high-quality medical images. 
\end{abstract}

\begin{IEEEkeywords}
Generative Adversarial Network, U-Net, Data Augmentation, Medical Image Analysis
\end{IEEEkeywords}

\section{Introduction}
\label{sec:introduction}
In recent decade, deep learning has attracted increasing research interests for the studies of medical imaging computing and its applications. One of the biggest challenges in applying deep learning to the medical imaging domain is to learn generalizable feature patterns from small datasets or a limited number of annotated samples. Deep learning based methods require a large amount of annotated training samples to support the inference which is hard to be fulfilled on medical imaging analysis~\cite{roth2015improving,litjens2017survey,zhang2019multi}. In medical imaging tasks, annotations are conducted by radiologists with expert knowledge about the data and the related tasks. Benefit from increasingly released medical datasets and grand challenges, the shortage of dataset is relieved to some extent. However, those datasets are still in limited size as they inevitably require laborious work from radiologists~\cite{frid2018gan}.

To overcome this problem, data augmentation has been popularly  utilized. The most commonly used data augmentation strategy is dataset manipulation including various simple modifications of the data, such as translation, rotation, flip, crop, and scale~\cite{chen2020residual,chen2020momentum}. These data augmentation methods have been widely applied to enrich the training set, thereby improving the model performance in various computer vision tasks~\cite{zhu2017unpaired,liu2017unsupervised}. Image modification can introduce some pixel-level side information to improve the performance. However, pixel-level modification cannot introduce new images but only the variants of the original one, and hence is still likely to suffer from the over-fitting problem. Instead, synthetic data augmentation method is considered to be more reasonable alternative as it can generate sophisticated types of data based on the original images. Generative adversarial networks (GAN) is a representative of the synthetic data augmentation method~\cite{goodfellow2014generative} and capable of providing more variability to enrich the dataset.

Inspired by the game theory, GAN aims to achieve the nash equilibrium~\cite{goodfellow2014generative} inside the model. GAN consists of two major networks which are trained jointly under an adversarial setting, where one network generates fake images based on inputs and the other network distinguishes the generated images from the real images. GANs have been increasingly applied to image synthesis~\cite{goodfellow2014generative,gulrajani2017improved}, such as denoising~\cite{yang2018low}, image translation~\cite{isola2017image}, etc. In addition, multiple variants of GANs were proposed to generate high quality realistic natural images~\cite{mirza2014conditional}, story visualization~\cite{li2019storygan}, and synthesize high-resolution images based on the low-resolution one~\cite{ledig2017photo}.

\begin{figure*}[ht]
    \centering
    \includegraphics[width=\linewidth]{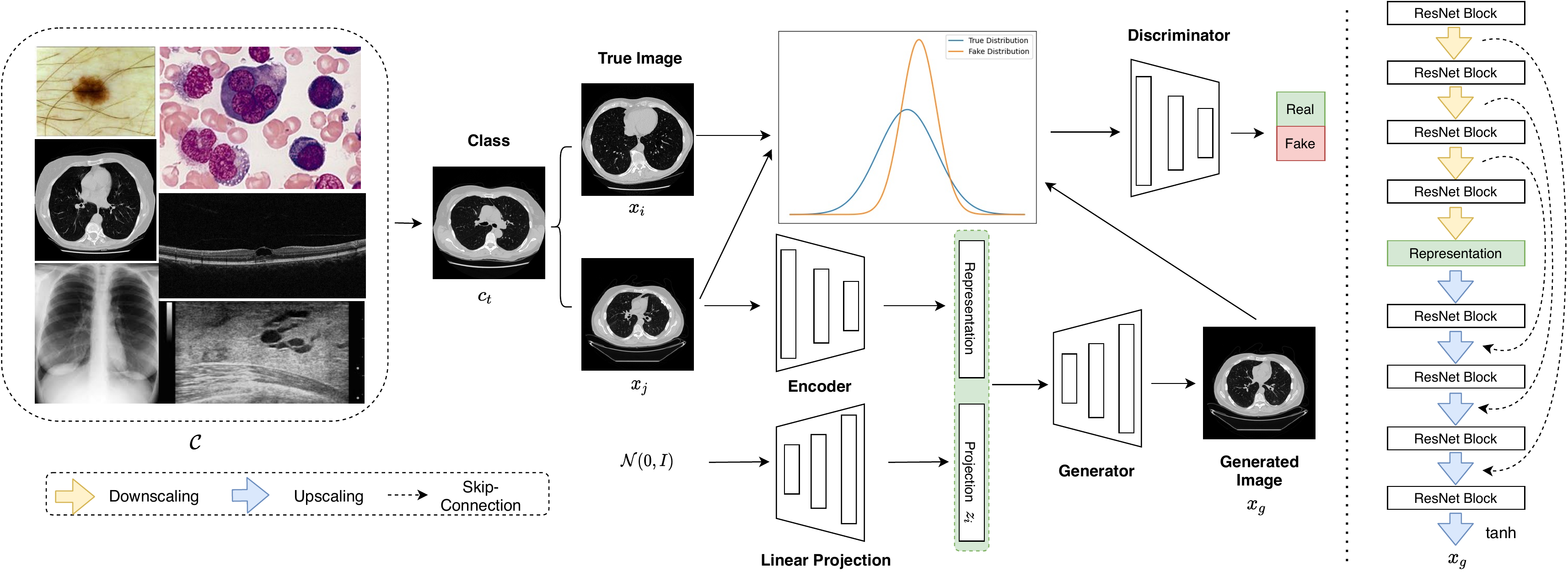}
    \caption{Structure of the proposed model. Given an arbitrary class $c_t \in \mathcal{C}$, our model can generate the corresponding images based on sampled image $x_j$. Two different samples will be sampled from the given class $c_t$ each time to support the discriminator. $x_j$ will be encoded into a latent representation and concatenated with the Gaussian variable as the final representation to be used for generation. The generated image $x_g$ will be fed into discriminator together with real data $x_i,x_j$. The discriminator is designed to distinguish two distributions, i.e., true distribution $\{x_i,x_j\}$ and fake distribution $\{x_i,x_g\}$. The generator aims to make those two distributions as similar as possible.}
    \label{fig:structure}
\end{figure*}

Recently, several studies on medical imaging adopted the GANs and their variants as the main framework~\cite{frid2018gan,zhang2019noise,nie2017medical}. Most studies have employed the GAN technique for generating images or medical cross-modality translations. Zhang et al.~\cite{zhang2019noise} applied GAN to reduce the intrinsic noises in multi-sourced dataset as different devices would generate different types of noises and such site effects would significantly affect data distribution. Zhang et al.~\cite{zhang2019skrgan} proposed SkrGAN by incorporating the global contextual information which is fine foreground structures to improve the image quality. Xue et al. ~\cite{xue2018segan} used two GANs to learn the relationship between brain MRI images and a brain tumor segmentation map. Fird et al.~\cite{frid2018gan} utilized GAN to generate liver lesion images to improve the CNN classification performance. Besides, GANs have also been successfully applied to segmentation. Dong et al.~\cite{dong2019neural} adopted GAN to do the neural architecture search to find the best way to make the segmentation for chest organs. Khosravan et al.~\cite{khosravan2019pan} introduced a projection module into GAN to boost the performance of segmentor on the lung.

However, most of the existing studies have been  focusing on a specific task or domain, and there is no robust method that is generalizable across various domains. In this study, we aim to design a domain-free GAN structure which is suitable for any domains instead of a specific one. Specifically, the proposed method can be used in any domain such as X-Ray, CT scan, pathology etc. In addition, vanilla GANs suffer from the training instability problem which is hard to make the model converge~\cite{arjovsky2017wasserstein}. We employ the Wasserstein GAN as the main framework in our model as it has shown higher training stability. U-Net is a well-known structure on medical imaging analysis, especially in segmentation~\cite{chen2020residual,zhou2018unet++}. Segmentation aims to find the pixel-level abnormality, which requires strong feature extraction capability. The generator in GANs requires a similar capability. Hence, we utilize U-Net as the generator in our study. U-Net is similar to auto-encoder which can learn the latent representation and reconstruct the output with the same size as the input. In order to fulfill the generation requirement, we concatenate a Gaussian variable into the latent representation to ensure that it will not generate the same image each time. The major contribution of  our study can be summarized as following:
The contribution of this paper can be summarized as following:
\begin{itemize}
    \item We propose a new variant of GAN named generative adversarial U-Net for the domain-free medical image augmentation. Images generated by the proposed method have a better quality than vanilla GANs and its well-known variant conditional GANs.
    \item To leverage the superior features extraction capability, we first dissemble U-Net into encoder and generator. Then, we assemble the generator into the GAN structure to generate images.
    \item Extensive experiments are conducted on eight different datasets in different domains include CT scan, Pathology, Chest X-Ray, Dermatoscope, Ultrasound, and Optical Coherence Tomography. Experimental results demonstrate high generalizability and robustness across various data domains.
\end{itemize}

\section{Methodology}
In this section, the proposed generative adversarial U-Net will be briefly introduced. We will start by describing the overall structure of the developed deep learning model followed by explaining the components including Residual U-Net generator, discriminator and the training strategy. The overall flowchart of our developed method is illustrated  in Fig.~\ref{fig:structure}.
\subsection{Overview}
U-Net was first proposed in~\cite{ronneberger2015u} and has been widely used in medical image segmentation~\cite{chen2020residual}. It is a type of artificial neural network by using the auto-encoder structure with skip connections. The encoder is designed to extract features from the given images and the decoder is to construct the segmentation map by using those extracted spatial features. The encoder follows a similar structure like fully convolutional networks (FCN)~\cite{long2015fully} with the stacked convolutional layers. To be specific, the encoder consists of a sequence of blocks for down-sampling operations, with each block including several convolution layers followed by max-pooling layers. The number of filters in the convolutional layers is doubled after each down-sampling operation. In the end, the encoder outputs a learned feature map for the input image.

Differently, the decoder is designed for up-sampling and constructing the image segmentation. The decoder first utilizes a  deconvolutional layer to up-sample the feature map generated by the encoder. The deconvolutional layer contains the transposed convolution operation and will halve the number of filters in the output. It is followed by a sequence of up-sampling blocks which consist of two convolution layers and a deconvolutional layer. Then, another convolutional layer is used as the final layer to generate the segmentation result. The final layer adopts Sigmoid function as the activation function while all other layers use ReLU function. 

In addition, the U-Net concatenates parts of the encoder features with the decoder, which is known as the skip-connection in ResNet~\cite{he2016deep}. For each block in the encoder, the result of the convolution before the max-pooling is transferred to the decoder symmetrically. In decoder, each block receives the feature representation learned from encoder, and concatenates them with the output of the deconvolutional layer. The concatenated result is then forwardly propagated to the consecutive block. This concatenation operation is useful for the decoder to capture the possible lost features by the max-pooling~\cite{zhou2018unet++}.

\subsection{U-Net Based Generative Model}
As mentioned previously, U-Net demonstrates state-of-the-art performance on medical image segmentation task,  showing its superiority on the medical image feature extraction. Hence, we utilize the U-Net as the main structure for the proposed generative model. GANs $G(z)$ are generative models that aim to learn a mapping from a random noise vector $z$ to the output image $x_g$, $G:z\rightarrow x_g$~\cite{goodfellow2014generative}.
\begin{align}
    z \in \mathcal{N}(0,I) \label{eq1}
\end{align}
However, the images generated by GANs are randomized which is hard to define the label. Hence, we use the conditional GANs~\cite{mirza2014conditional} instead. Conditional GANs learn a mapping from a random noise vector $z$ and observed images $x_i$ for class $c_{t}$ to the output image $x_g$, $G_c:(z,x_i)\rightarrow x_g$. Generator $G$ is trained to generate images that cannot be distinguished from the ``real'' images by an adversarial trained discriminator $D$. Discriminator $D$ is trained to detect fake images produced by generator. Alternatively, GANs are designed to conduct the distribution discrepancy measurement between the generated data and real data.
\begin{table*}[ht]
    \centering
    \caption{Summary description for the datasets used in our experimental study.}
    \begin{tabular}{c|c|c|c|c|c}
         \hline
         Name &  Modality & Tasks & \# Training & \# Validation & \# Test\\ \hline 
         NCT-CRC-HE-100K & Pathology & Multi-Class(9) & 70,000 & 10,000 & 20,000 \\ 
         ChestXray8 & Chest X-ray & Multi-Class(14) & 78,468 & 11,219 & 22,433 \\
         HAM10000 & Dermatoscope & Multi-Class(7) & 7,007 &  1,003 & 2,005 \\
         OCT 2017 &  OCT & Multi-Class(4) & 76,516 & 10,930 & 21,863 \\ 
         X-Ray OCT 2017 & Chest X-Ray & Binary(2) & 4,099 & 386 & 1,171 \\
         LUNA & Lung CT & Multi-Class(3) & 622 & 88 & 178 \\
         BreastUltra & Breast Ultrasound & Binary(2) & 546 & 78 & 156\\
         LiTS & Abdominal CT & Multi-Class(11) &  41,195 & 5,885 & 11,770\\
         \hline
    \end{tabular}
    \label{tab:stat}
\end{table*}
The objective function of the conditional GANs can be expressed as:
\begin{align}
    \mathcal{L}_{c}(G_{c},D) = & \mathbb{E}_{x_{i},x_{g}}[\log D(x_{i},x_{g})] \nonumber\\ 
    & + \mathbb{E}_{x_{i},z}[\log (1-D(x_{i},G_{c}(x_{i},z)))] \label{eq2}
\end{align}
where $G_c$ tries to minimize the objective function against $D$ that tries to maximize it. Mathematically, it is formulated as follows:
\begin{align}
    G = \arg \min_{G_c}\max_D \mathcal{L}_c(G_c,D) \label{eq3}
\end{align}
where $G$ is the result generator when Eq.\eqref{eq3} reaches the Nash equilibrium. 
However, conditional GANs have similar limitations with the traditional GANs, which suffers from training instability and model collapse problems. Hence, we use the Wasserstein GAN~\cite{arjovsky2017wasserstein} as the main structure for our generative model. To be specific, a normal GAN minimizes JS-Divergence which is shown on Eq.\eqref{eq2}, whereas objective function for Wasserstein GAN is:
\begin{align}
    \mathcal{L}_w(G_w,D_w) = &\mathbb{E}_{x_i,z}[(D_w(x_i,G_w(x_i,z)))]\nonumber\\ 
    & - \mathbb{E}_{x_i,x_g}[D_w(x_i,x_g)]  \label{eq4}
\end{align}
Furthermore, an gradient penalty is introduced~\cite{gulrajani2017improved} to enforce the Lipschitz constraint for Wasserstein GAN which is:
\begin{align}
    \lambda \mathbb{E}_{\hat{x}\sim \mathbb{P}_{\hat{x}}}[(\|\nabla_{\hat{x}}D(\hat{x})\|_2 -1)^2]  \label{eq5}
\end{align}
where the $\mathbb{P}_{\hat{x}}$ sampling uniformly along straight lines between pairs of points sampled from the data distribution and generator distribution. The $\hat{x}$ is the combination of the original images and generated ones sampled from an uniform distribution $\mathcal{U}(0,1)$ with a control factor $\epsilon$. 

A generative adversarial network could be used to conduct data augmentation. Given a certain class $c_t$ and corresponding data point $x$, we are able to learn a representation of the input image $r_x$ through the encoder such that $r_x = g(x)$ where $g(\cdot)$ represents the encoder network. In addition, a latent Gaussian variable $z_i$ is introduced into the learned representation to provide the variation with the following form
\begin{align}
    z_i \in g_l(\mathcal{N}(0,I))
\end{align}
where $g_l$ is the linear projection to project the Gaussian noise into vector form so that it can be concatenated with the learned representation. Once the representation is learned, it will be fed into the generator to generate images. In the proposed method, U-Net is split up into encoder and generator. The structure of U-Net can be found in the right side of Fig.~\ref{fig:structure}. The ResNet block is used as the basic unit, which is defined as
\begin{align}
    F(k) = \sum_{j=1} w_jk_j + r_k
\end{align}
where $k_j$ is $k-th$ layer and $w_j$ is the corresponding trainable weight and $r_k$ is the residual. In addition, different from the traditional U-Net, we use leaky ReLU $f(x)$ as the activation function.
\begin{align}
    f(x)=\begin{cases} 
      0.01x & x\leq 0 \\
      x &  x> 0 
   \end{cases}
\end{align}
It is  worth noting that the generated image $x_g$ and original image $x_j$ are both provided to the discriminator. We want to ensure that the generator is capable of generating the image that is related to but different from the original image $x_j$. That is, the generated image $x_g$ is supposed to be drawn from the same class as $x_j$ other than just a duplicate or simple modification of $x_j$. By providing the current image $x_j$, we can prevent generator from simply encoding it. In addition, the class information is provided where the generator can better learn the generalized pattern across all classes. 
\subsection{Model Architectures}
The generator contains eight blocks where each block has four $3\times 3$ convolutional layers with batch renormalization~\cite{ioffe2017batch} followed by a downscaling or upscaling layer. Downscaling layers are convolutions with stride 2 followed by leaky ReLU, batch normalization and dropout. Upscaling layers are deconvolution with stride $\frac{1}{2}$ followed by leaky ReLU, batch renormalization and dropout. As aforementioned, we also maintain the skip-connections inside the generator. We use the similar strategy with ResNet, we use a $1\times 1$ convolutional layer to pass features between blocks. DenseNet~\cite{huang2017densely} is adopted as the discriminator. Layer normalization is applied instead of batch normalization as we find it has a better performance. Discriminator contains four dense blocks and four transition layers where each dense block contains four convolutional layers and ends with a dropout layer. The reason why we apply the dropout at the last layer is that it can avoid the overfitting.
\subsection{Optimization}
To optimize our networks, we follow the standard approach introduced in~\cite{goodfellow2014generative}: alternately update one gradient descent step on $D$, then one step on $G$. As suggested in the original WGAN paper, we train the model based on the algorithm~\ref{alg:d}.
\begin{algorithm}[ht]
\SetAlgoLined
 \SetKwInOut{Input}{input}
 \Input{gradient penalty coefficient $\lambda$, Adam parameter $\alpha,\beta_1,\beta_2$, batch size $m$, input image $x_i$}
 \Input{discriminator parameter $w_0$, generator parameter $\theta_0$}
 \While{$\theta$ not converged}{
    \For{$t=1,\cdots$}{
        \For{$i=1,\cdots,m$}{
            Sample real data $x_j$, random noise from Eq.\eqref{eq1}, random number $\epsilon\sim \mathcal{U}(0,1)$\;
            $x_g = G_\theta(z,x_j)$\;
            $\hat{x} = \epsilon x_i + (1-\epsilon) x_g$\;
            $L^i = $Eq.\eqref{eq4} + Eq.\eqref{eq5}\;    
        }
        $w \leftarrow$ Adam($\nabla_w \frac{1}{m}\sum_{i=1}^m L^i,w,\alpha,\beta_1,\beta_2$)\;
    }
    Sample batch of latent variable $\{z_i\}_{i=1}^m \sim \mathcal{N}(0,I)$\;
    $\theta\leftarrow$ Adam($\nabla_\theta \frac{1}{m}\sum_{i=1}^m -D_w(x_g)),\theta,\alpha,\beta_1,\beta_2$)\;
 }
 \caption{Training algorithm for our model}
 \label{alg:d}
\end{algorithm}
\begin{table*}[ht]
    \centering
    \caption{Quantitative comparison of the quality of generated images with 95\% confidence interval.}
    \begin{tabular}{c|cc|cc|cc|cc}
     \hline 
      & \multicolumn{2}{c|}{NCT-CRC-HE} & \multicolumn{2}{c|}{ChestXray8} & \multicolumn{2}{c|}{HAM10000} & \multicolumn{2}{c}{OCT 2017}  \\ \hline
      & FID$\downarrow$ & PA$\uparrow$ & FID & PA & FID & PA & FID & PA \\ \hline
      GAN &  41.65 $\pm$ 0.12 & 0.52 $\pm$ 0.11 & 40.02 $\pm$ 0.22 & 0.51 $\pm$ 0.09 & 47.02 $\pm$ 0.12 & 0.40 $\pm$ 0.10 & 42.12 $\pm$ 0.18 & 0.51 $\pm$ 0.08\\
      CGAN & 32.65 $\pm$ 0.05 & 0.65 $\pm$ 0.08 & 30.88 $\pm$ 0.09 & 0.59 $\pm$ 0.05 & 45.52 $\pm$ 0.11 & 0.44 $\pm$ 0.09 & 37.72 $\pm$ 0.15 & 0.58 $\pm$ 0.06\\ 
      Ours & \textbf{29.65 $\pm$ 0.06} & \textbf{0.68 $\pm$ 0.07} & \textbf{20.78 $\pm$ 0.07} & \textbf{0.63 $\pm$ 0.06} & \textbf{42.22 $\pm$ 0.10} & \textbf{0.50 $\pm$ 0.06} & \textbf{34.22 $\pm$ 0.12} & \textbf{0.61 $\pm$ 0.04}\\\hline
    \end{tabular}
    
    \begin{tabular}{c|cc|cc|cc|cc}
     \hline 
      & \multicolumn{2}{c|}{X-Ray OCT} & \multicolumn{2}{c|}{LUNA} & \multicolumn{2}{c|}{BreastUltra} & \multicolumn{2}{c}{LiTS} \\ \hline
      & FID$\downarrow$ & PA$\uparrow$ & FID & PA & FID & PA & FID & PA \\ \hline
      GAN & 47.02 $\pm$ 0.12 & 0.40 $\pm$ 0.10 & 62.65 $\pm$ 0.22 & 0.33 $\pm$ 0.08 & 63.02 $\pm$ 0.12 & 0.35 $\pm$ 0.10 & 43.12 $\pm$ 0.13 & 0.51 $\pm$ 0.07\\
      CGAN & 45.32 $\pm$ 0.11 & 0.43 $\pm$ 0.09 & 60.15 $\pm$ 0.18 & 0.37 $\pm$ 0.07 & 61.62 $\pm$ 0.11 & 0.41 $\pm$ 0.09 & 42.01 $\pm$ 0.11 & 0.53 $\pm$ 0.09 \\
      Ours & \textbf{41.88 $\pm$ 0.10} & \textbf{0.52 $\pm$ 0.08} & \textbf{52.15 $\pm$ 0.13} & \textbf{0.49 $\pm$ 0.08} & \textbf{55.62 $\pm$ 0.09} & \textbf{0.48 $\pm$ 0.07} & \textbf{37.98 $\pm$ 0.14} & \textbf{0.59 $\pm$ 0.07}\\\hline
    \end{tabular}
    \label{tab:gan_result}
\end{table*}

\begin{table*}[ht]
    \centering
    \caption{Classification results for top-5 with 95\% confidence interval with(W) and without(W/O) augmentation}
    \begin{tabular}{>{\rowmac}c>{\rowmac}c|>{\rowmac}c>{\rowmac}c>{\rowmac}c>{\rowmac}c|>{\rowmac}c>{\rowmac}c>{\rowmac}c>{\rowmac}c<{\clearrow}}
     \hline 
      \multirow{2}{*}{} & & \multicolumn{4}{c|}{NCT-CRC-HE} & \multicolumn{4}{c}{ChestXray8}  \\ \cline{3-10}
      & & Acc & Pre & Rec & AUC & Acc & Pre & Rec & AUC \\ \hline
      \multirow{2}{*}{ResNet18} &  W & \setrow{\bfseries} 0.889 $\pm$ 0.07 &  0.859 $\pm$ 0.08 & 0.866 $\pm$ 0.09 & 0.980 $\pm$ 0.10 &  0.942 $\pm$ 0.10 & 0.920 $\pm$ 0.18 & 0.930 $\pm$ 0.16 & 0.970 $\pm$ 0.11\\  
      &  W/O&  0.806 $\pm$ 0.09 & 0.783 $\pm$ 0.11 & 0.792 $\pm$ 0.14 & 0.945 $\pm$ 0.12 & 0.921 $\pm$ 0.11 & 0.908 $\pm$ 0.13 & 0.916 $\pm$ 0.13 & 0.958 $\pm$ 0.16\\ \hline
      \multirow{2}{*}{ResNet50} &  W & \setrow{\bfseries} 0.892 $\pm$ 0.08 & 0.831 $\pm$ 0.06 & 0.866 $\pm$ 0.07 & 0.977 $\pm$ 0.05 & 0.959 $\pm$ 0.08 & 0.930 $\pm$ 0.07 & 0.941 $\pm$ 0.06 & 0.979 $\pm$ 0.06\\  
      &  W/O & 0.846 $\pm$ 0.09 & 0.820 $\pm$ 0.09 & 0.831 $\pm$ 0.11 & 0.969 $\pm$ 0.08 & 0.947 $\pm$ 0.07 & 0.922 $\pm$ 0.08 & 0.931 $\pm$ 0.09 & 0.921 $\pm$ 0.09\\ \hline
      \multirow{2}{*}{DenseNet161} &  W & \setrow{\bfseries} 0.888 $\pm$ 0.07 & 0.825 $\pm$ 0.08 & 0.860 $\pm$ 0.09 & 0.974 $\pm$ 0.07 & 0.949 $\pm$ 0.05 & 0.932 $\pm$ 0.06 & 0.940 $\pm$ 0.06 & 0.971 $\pm$ 0.07\\  
      & W/0 & 0.840 $\pm$ 0.09 & 0.812 $\pm$ 0.10 & 0.821 $\pm$ 0.10 & 0.958 $\pm$ 0.08 & 0.927 $\pm$ 0.09 & 0.902 $\pm$ 0.08 & 0.912 $\pm$ 0.09 & 0.910 $\pm$ 0.09\\ \hline
      \multirow{2}{*}{SVM} & W & \setrow{\bfseries} 0.791 $\pm$ 0.18 & 0.768 $\pm$ 0.17 & 0.770 $\pm$ 0.18 & 0.891 $\pm$ 0.12 & 0.823 $\pm$ 0.20 & 0.802 $\pm$ 0.19 & 0.814 $\pm$ 0.18 & 0.900 $\pm$ 0.19\\  
      &  W/O &  0.721 $\pm$ 0.17 & 0.708 $\pm$ 0.16 & 0.710 $\pm$ 0.18 & 0.802 $\pm$ 0.13 & 0.800 $\pm$ 0.19 & 0.782 $\pm$ 0.16 & 0.790 $\pm$ 0.15 & 0.811 $\pm$ 0.13 \\ \hline
    \end{tabular}

    \begin{tabular}{>{\rowmac}c>{\rowmac}c|>{\rowmac}c>{\rowmac}c>{\rowmac}c>{\rowmac}c|>{\rowmac}c>{\rowmac}c>{\rowmac}c>{\rowmac}c<{\clearrow}}
     \hline 
      \multirow{2}{*}{} & & \multicolumn{4}{c|}{HAM10000} & \multicolumn{4}{c}{OCT 2017}  \\ \cline{3-10}
      & & Acc & Pre & Rec & AUC & Acc & Pre & Rec & AUC \\ \hline
      \multirow{2}{*}{ResNet18} &  W & \setrow{\bfseries} 0.802 $\pm$ 0.07 & 0.788 $\pm$ 0.06 & 0.790 $\pm$ 0.07 & 0.932 $\pm$ 0.05 & 0.810 $\pm$ 0.07 & 0.775 $\pm$ 0.06 & 0.781 $\pm$ 0.07 & 0.940 $\pm$ 0.05\\  
      &  W/O&  0.750 $\pm$ 0.09 & 0.722 $\pm$ 0.08 & 0.731 $\pm$ 0.07 & 0.911 $\pm$ 0.06 & 0.757 $\pm$ 0.07 & 0.729 $\pm$ 0.07 & 0.740 $\pm$ 0.06 & 0.904 $\pm$ 0.07\\ \hline
      \multirow{2}{*}{ResNet50} &  W & \setrow{\bfseries} 0.800 $\pm$ 0.06 & 0.777 $\pm$ 0.07 & 0.781 $\pm$ 0.05 & 0.930 $\pm$ 0.06 & 0.801 $\pm$ 0.08 & 0.769 $\pm$ 0.07 & 0.775 $\pm$ 0.06 & 0.937 $\pm$ 0.05\\
      & W/O &  0.744 $\pm$ 0.06 & 0.723 $\pm$ 0.07 & 0.732 $\pm$ 0.05 & 0.910 $\pm$ 0.06 & 0.751 $\pm$ 0.07 & 0.737 $\pm$ 0.06& 0.734 $\pm$ 0.05 & 0.902 $\pm$ 0.05\\ \hline
      \multirow{2}{*}{DenseNet161} &  W & \setrow{\bfseries} 0.801 $\pm$ 0.08 & 0.775 $\pm$ 0.06 & 0.779 $\pm$ 0.07 & 0.925 $\pm$ 0.07 & 0.799 $\pm$ 0.07 & 0.753 $\pm$ 0.05 & 0.761$\pm$ 0.06 & 0.929$\pm$ 0.06\\  
      & W/O&  0.739 $\pm$ 0.06 & 0.718 $\pm$ 0.09 & 0.721 $\pm$ 0.08 & 0.899 $\pm$ 0.05 & 0.741 $\pm$ 0.07 & 0.703 $\pm$ 0.06 & 0.714 $\pm$ 0.07 & 0.890 $\pm$ 0.05\\ \hline
      \multirow{2}{*}{SVM} & W & \setrow{\bfseries} 0.701 $\pm$ 0.11 & 0.681 $\pm$ 0.10 & 0.690 $\pm$ 0.11 & 0.887 $\pm$ 0.09 & 0.732 $\pm$ 0.12 & 0.700 $\pm$ 0.09 & 0.711 $\pm$ 0.10 & 0.886 $\pm$ 0.08\\  
      &  W/O &  0.678 $\pm$ 0.08 & 0.650 $\pm$ 0.09  & 0.660 $\pm$ 0.09  & 0.867 $\pm$ 0.08 & 0.699 $\pm$ 0.10 & 0.671 $\pm$ 0.07 & 0.679 $\pm$ 0.08 & 0.876 $\pm$ 0.06 \\ \hline
    \end{tabular}
    
    \begin{tabular}{>{\rowmac}c>{\rowmac}c|>{\rowmac}c>{\rowmac}c>{\rowmac}c>{\rowmac}c|>{\rowmac}c>{\rowmac}c>{\rowmac}c>{\rowmac}c<{\clearrow}}
     \hline 
      \multirow{2}{*}{} & & \multicolumn{4}{c|}{X-Ray OCT} & \multicolumn{4}{c}{LUNA}  \\ \cline{3-10}
      & & Acc & Pre & Rec & AUC & Acc & Pre & Rec & AUC \\ \hline
      \multirow{2}{*}{ResNet18} & W & \setrow{\bfseries} 0.877 $\pm$ 0.08 & 0.840 $\pm$ 0.06 & 0.851 $\pm$ 0.08 & 0.961 $\pm$ 0.08 & 0.852 $\pm$ 0.06 & 0.801 $\pm$ 0.06 & 0.821$\pm$ 0.05 & 0.944 $\pm$ 0.06 \\  
                                & W/O & 0.843 $\pm$ 0.09  & 0.821 $\pm$ 0.07 & 0.833 $\pm$ 0.08 & 0.957 $\pm$ 0.07 & 0.822 $\pm$ 0.08 & 0.781 $\pm$ 0.06 & 0.791 $\pm$ 0.09 & 0.919 $\pm$ 0.08 \\ \hline
      \multirow{2}{*}{ResNet50} & W & \setrow{\bfseries} 0.880 $\pm$ 0.09 & 0.849 $\pm$ 0.07 & 0.858 $\pm$ 0.08 & 0.974 $\pm$ 0.06 &                              0.850 $\pm$ 0.08 & 0.799 $\pm$ 0.07 & 0.818 $\pm$ 0.09 & 0.942 $\pm$ 0.05\\  
                                & W/O & 0.861 $\pm$ 0.08 & 0.836 $\pm$ 0.07 & 0.844 $\pm$ 0.08 & 0.970 $\pm$ 0.07 & 0.817 $\pm$ 0.09 & 0.776 $\pm$ 0.08 & 0.780 $\pm$ 0.06 & 0.916 $\pm$ 0.05 \\ \hline
    \multirow{2}{*}{DenseNet161}& W & \setrow{\bfseries} 0.866 $\pm$ 0.08 & 0.831 $\pm$ 0.06 & 0.850 $\pm$ 0.05 & 0.951 $\pm$ 0.07 &                                0.852 $\pm$ 0.08 & 0.797 $\pm$ 0.06 & 0.809 $\pm$ 0.05 & 0.939 $\pm$ 0.05\\  
                                & W/O & 0.857 $\pm$ 0.09 & 0.819 $\pm$ 0.08 & 0.830 $\pm$ 0.08 & 0.949 $\pm$ 0.08 & 0.814 $\pm$ 0.10 & 0.770 $\pm$ 0.07 & 0.777 $\pm$ 0.08 & 0.913 $\pm$ 0.09\\ \hline
      \multirow{2}{*}{SVM}      & W & \setrow{\bfseries} 0.821 $\pm$ 0.11 & 0.799 $\pm$ 0.09 & 0.801 $\pm$ 0.09 & 0.930 $\pm$ 0.08 &                              0.822 $\pm$ 0.11 & 0.779 $\pm$ 0.08 & 0.791 $\pm$ 0.07 & 0.900 $\pm$ 0.08\\  
                                & W/O  & 0.800 $\pm$ 0.14 & 0.754 $\pm$ 0.11 & 0.769 $\pm$ 0.12 & 0.929 $\pm$ 0.10 & 0.788 $\pm$ 0.12 & 0.742 $\pm$ 0.10 & 0.750 $\pm$ 0.10 & 0.899 $\pm$ 0.09\\ \hline
    \end{tabular}
    
    \begin{tabular}{>{\rowmac}c>{\rowmac}c|>{\rowmac}c>{\rowmac}c>{\rowmac}c>{\rowmac}c|>{\rowmac}c>{\rowmac}c>{\rowmac}c>{\rowmac}c<{\clearrow}}
     \hline 
      \multirow{2}{*}{} & & \multicolumn{4}{c|}{BreastUltra} & \multicolumn{4}{c}{LiTS}  \\ \cline{3-10}
      & & Acc & Pre & Rec & AUC & Acc & Pre & Rec & AUC \\ \hline
      \multirow{2}{*}{ResNet18} & W & \setrow{\bfseries} 0.889 $\pm$ 0.07 & 0.842 $\pm$ 0.08 & 0.851 $\pm$ 0.07 & 0.909 $\pm$ 0.06 &                              0.944 $\pm$ 0.08 & 0.899 $\pm$ 0.06 & 0.909 $\pm$ 0.06 & 0.981 $\pm$ 0.07\\  
                                & W/O   & 0.859 $\pm$ 0.08 & 0.820 $\pm$ 0.08 & 0.831 $\pm$ 0.07 & 0.897 $\pm$ 0.06 & 0.921 $\pm$ 0.08 & 0.881 $\pm$ 0.08 & 0.892 $\pm$ 0.06 & 0.979 $\pm$ 0.06\\ \hline
      \multirow{2}{*}{ResNet50} & W & \setrow{\bfseries} 0.900 $\pm$ 0.08 & 0.864 $\pm$ 0.07 & 0.871 $\pm$ 0.08 & 0.919 $\pm$ 0.06 &                              0.941 $\pm$ 0.07 & 0.907 $\pm$ 0.06 & 0.915 $\pm$ 0.07 & 0.979 $\pm$ 0.06\\  
                                & W/O   & 0.878 $\pm$ 0.09 & 0.843 $\pm$ 0.08 & 0.850 $\pm$ 0.09 & 0.915 $\pm$ 0.08 & 0.923 $\pm$ 0.08 & 0.902 $\pm$ 0.05 & 0.911 $\pm$ 0.05 & 0.978 $\pm$ 0.06 \\ \hline
    \multirow{2}{*}{DenseNet161}& W & \setrow{\bfseries} 0.884 $\pm$ 0.09 & 0.839 $\pm$ 0.09 & 0.845 $\pm$ 0.09 & 0.907 $\pm$ 0.08 &                                0.939 $\pm$ 0.08 & 0.890 $\pm$ 0.08 & 0.905 $\pm$ 0.07 & 0.979 $\pm$ 0.08\\  
                                & W/O   & 0.853 $\pm$ 0.08 & 0.821 $\pm$ 0.07 & 0.840 $\pm$ 0.08 & 0.879 $\pm$ 0.07 & 0.915 $\pm$ 0.07 & 0.889 $\pm$ 0.07 & 0.900 $\pm$ 0.06 & 0.976 $\pm$ 0.06\\ \hline
      \multirow{2}{*}{SVM}      & W & \setrow{\bfseries} 0.833 $\pm$ 0.11 & 0.770 $\pm$ 0.12 & 0.781 $\pm$ 0.09 & 0.838 $\pm$ 0.08 &                              0.899 $\pm$ 0.10& 0.859 $\pm$ 0.08 & 0.870 $\pm$ 0.09 & 0.955 $\pm$ 0.08 \\  
                                & W/O   & 0.810 $\pm$ 0.14 & 0.767 $\pm$ 0.13 & 0.779 $\pm$ 0.11 & 0.833 $\pm$ 0.12 & 0.889 $\pm$ 0.10 & 0.854 $\pm$ 0.11 & 0.866 $\pm$ 0.10 & 0.951 $\pm$ 0.09\\ \hline
    \end{tabular}
    \label{tab:class_result}
\end{table*}
\section{Experiment}
\subsection{Experiments Setup}
The experiments are conducted on multiple public available datasets which are:
\begin{itemize}
    \item NCT-CRC-HE-100K~\cite{kather2019predicting}: A dataset contains $100,000$ non-overlapping image patches from hematoxylin and eosin stained histological images. Nine different types of tissues are involved.
    \item ChestXray8~\cite{wang2017chestx}: A dataset contains $112, 120$ frontal-view X-ray images of $30, 805$ unique patients with 14 disease image labels.
    \item HAM10000~\cite{tschandl2018ham10000}: A dataset contains $10,015$ multi-source dermatoscopic images of common
    pigmented skin lesions which belongs to seven different categories.
    \item Optical Coherence Tomography 2017 (OCT2017)~\cite{kermany2018identifying}: A dataset contains $109, 309$ optical coherence tomography images for four different type retinal diseases.
    \item X-Ray OCT 2017~\cite{kermany2018identifying}: A dataset extended from OCT2017 which contains $5, 856$ chest CT images from pneumonia and normal patients.
    \item LUNA~\cite{setio2017validation}: LUng Nodule Analysis is an open dataset consisting of 888 CT scans with three different classes: non-nodule, nodule $<$ 3 mm, and nodules $\geq$ 3 mm.
    \item BreastUltra~\cite{yap2017automated}: A dataset contains 780 breast ultrasound images belonging to three different classes: normal, benign, and malignant.
    \item LiTS~\cite{bilic2019liver}: Liver Tumor Segmentation Benchmark dataset which contains 3D computed tomography (CT). We use the similar method in~\cite{xu2019efficient} to transfer them into 2D images with the axial view.
\end{itemize}
OpenCV2 is utilized to resize all those medical images into $256\times 256\times 3$. We use the ratio 7:1:2 to split the datasets into training set, validation set and test set on patient level. The information of all these datasets is summarized on Table~\ref{tab:stat}.

The experiments were conducted on a machine with eight GPUs which includes six NVIDIA TITAN X Pascal GPUs and two NVIDIA TITAN RTXs. The model is implemented in Tensorflow.

Quality evaluation of the generated images is a challenging problem~\cite{salimans2016improved}. Traditional metrics, such as per-pixel mean square error, is hard to reflect the performance. Hence, we use the Fréchet Inception Distance (FID)~\cite{heusel2017gans} to measure the distance between the generated distribution and the real distribution. Lower FID indicates that the generated images have higher quality. In addition, we also use the Per-pixel Accuracy (PA) to measure the discriminability of the generated images~\cite{salimans2016improved,wang2016generative,isola2017image}.
In order to demonstrate the superiority of the proposed method, we select the following baselines to compare the quality.
\begin{itemize}
    \item Vanilla GAN~\cite{goodfellow2014generative}: the original version of the GAN.
    \item Conditional GAN~\cite{mirza2014conditional}: conditional GAN which consider about the label information.
\end{itemize}
It is worth to mention that U-Net is type of auto-encoder, but auto-encoders can not be used as baselines. Auto-encoders are widely used on reconstruction task instead of augmentation, it is not suitable in this manuscript.
For classification, we use several different metrics which are: 
\begin{itemize}
    \item Accuracy, which measures the percentage of correctly classified samples over the whole dataset
    \item Precision, which measures the percentage of true positives (TP) over all predicted positive samples
    \item Recall, used to measure the percentage of TPs over all positive samples
    \item Area-under-the-curve (AUC) which measures the relation between FPs and TPs.
\end{itemize}

The following baselines are selected as the classifier: ResNet-18, ResNet-50, DenseNet-161 and Support Vector Machine (SVM).
{
\begin{figure*}[ht]
    \centering
    \includegraphics[width=0.95\linewidth]{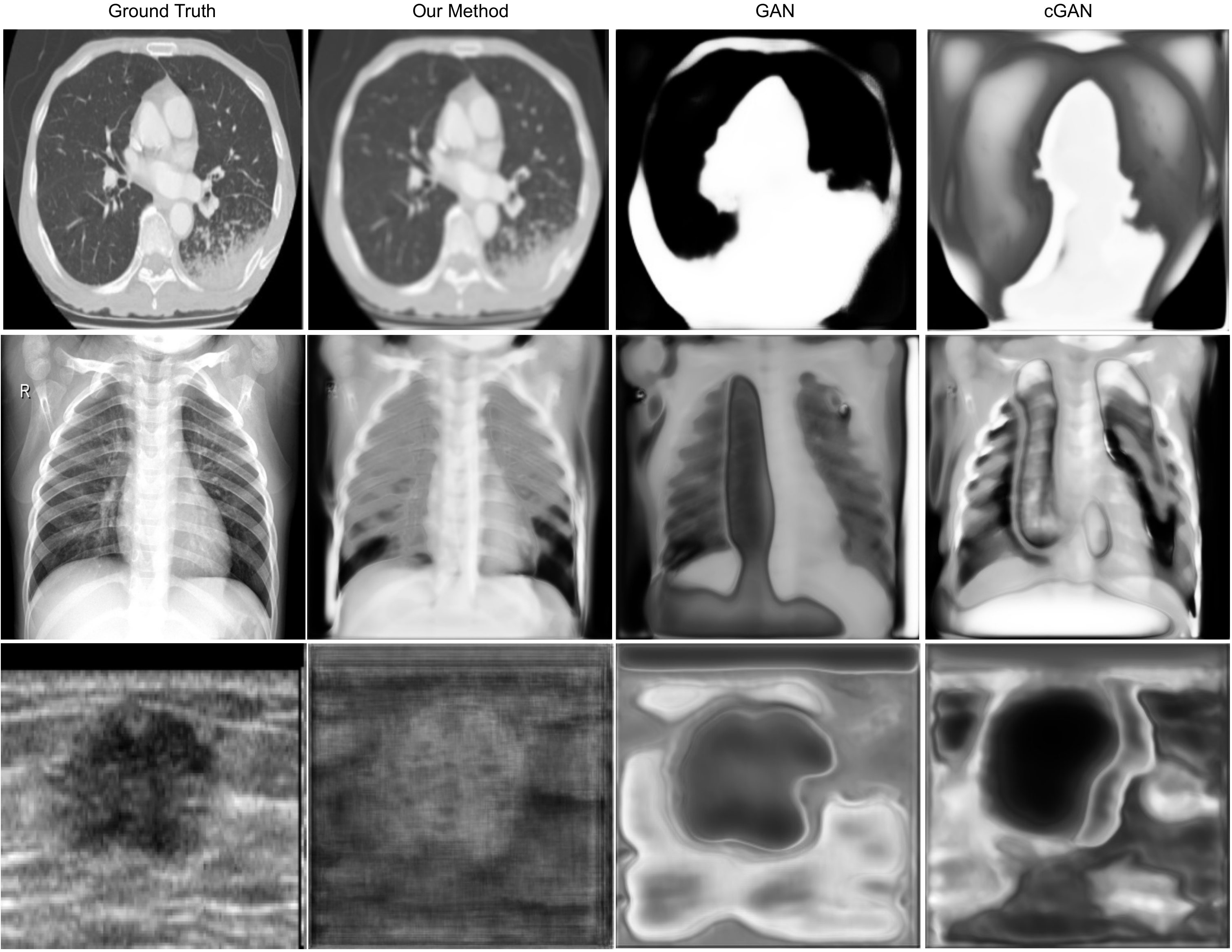}
    \caption{Generated images on three different domains: Lung CT, Chest X-Ray and Ultrasound. From left to right are: original images, our method, GAN generated image, and conditional GAN generated image. It is obvious that the images generated by our method is more similar with the original ones.}
    \label{fig:vis}
\end{figure*}
}
\subsection{Hyper Parameters Setting}
In this part, we briefly present the parameters setting used in our experiments. The gradient penalty coefficient $\lambda$ is 10. For Adma optimizer~\cite{kingma2014adam}, we set $\alpha = 0.0001,\beta_1=0,\beta_2=0.9$. The growth rate for Dense Block $k=64$, the number of training epoch is $5,000$. The classifiers are trained for $5,000$ epoches with Adma optimizer with $\alpha=0.001, \beta_1=0.9, \beta_2=0.99$.
\subsection{Results}
We first compared the performance of image generation among GAN, cGAN, and our method. The results summarized in Table~\ref{tab:gan_result} show that our method achieves the best result compared with GAN and conditional GAN on both metrics. In order to better analyze the generated images, we provide the visualization for those generated images. We use three different datasets: Luna, ChestXray8 and BreatUltra as the demo datasets. The visualization can be found on  Fig~\ref{fig:vis}. It's obviously that the image generated by our model are more similar with the ground truth.

We also provide the classification results with and without augmentation on all the eight datasets with four different classifiers on Table~\ref{tab:class_result}. We found that the performance of all the classifiers improved significantly after augmentation.

\section{Conclusion}
The lack of annotated medical images leads to a significant challenge for imaging-based medical studies, such as quick diagnosis, disease prediction, etc. Data augmentation is a popular approach to relieve this problem. In this paper, we propose a new method named generative adversarial U-Net for medical image augmentation. It can be used to generate multi-modalities data to relieve the data shortage which is commonly faced in medical imaging research. Specifically, we adjust the structure of the generative adversarial network to fit into the U-Net. We conduct extensive experiments on eight datasets with different modalities from binary classification to multi-class classification. Our experimental results demonstrated the superior performance of the proposed method over the state-of-the-art approaches on all of these datasets. In the future, we plan to extend our work into the more challenging few-shot learning scenario or semi-supervised learning scenario~\cite{zhang2019adversarial,yao2016learning} in which only a few samples even no samples available for certain class. We plan to investigate the transfer learning to help with enriching the current model to augment the ability of dealing with unseen samples from a brand new class.
\bibliographystyle{IEEEtran}
\bibliography{IEEEabrv,sample}

\begin{thebibliography}{10}
\providecommand{\url}[1]{#1}
\csname url@samestyle\endcsname
\providecommand{\newblock}{\relax}
\providecommand{\bibinfo}[2]{#2}
\providecommand{\BIBentrySTDinterwordspacing}{\spaceskip=0pt\relax}
\providecommand{\BIBentryALTinterwordstretchfactor}{4}
\providecommand{\BIBentryALTinterwordspacing}{\spaceskip=\fontdimen2\font plus
\BIBentryALTinterwordstretchfactor\fontdimen3\font minus
  \fontdimen4\font\relax}
\providecommand{\BIBforeignlanguage}[2]{{%
\expandafter\ifx\csname l@#1\endcsname\relax
\typeout{** WARNING: IEEEtran.bst: No hyphenation pattern has been}%
\typeout{** loaded for the language `#1'. Using the pattern for}%
\typeout{** the default language instead.}%
\else
\language=\csname l@#1\endcsname
\fi
#2}}
\providecommand{\BIBdecl}{\relax}
\BIBdecl

\bibitem{roth2015improving}
H.~R. Roth, L.~Lu, J.~Liu, J.~Yao, A.~Seff, K.~Cherry, L.~Kim, and R.~M.
  Summers, ``Improving computer-aided detection using convolutional neural
  networks and random view aggregation,'' \emph{IEEE transactions on medical
  imaging}, vol.~35, no.~5, pp. 1170--1181, 2015.

\bibitem{litjens2017survey}
G.~Litjens, T.~Kooi, B.~E. Bejnordi, A.~A.~A. Setio, F.~Ciompi, M.~Ghafoorian,
  J.~A. Van Der~Laak, B.~Van~Ginneken, and C.~I. S{\'a}nchez, ``A survey on
  deep learning in medical image analysis,'' \emph{Medical image analysis},
  vol.~42, pp. 60--88, 2017.

\bibitem{zhang2019multi}
X.~Zhang, X.~Chen, M.~Dong, H.~Liu, C.~Ge, and L.~Yao, ``Multi-task generative
  adversarial learning on geometrical shape reconstruction from eeg brain
  signals,'' \emph{arXiv preprint arXiv:1907.13351}, 2019.

\bibitem{frid2018gan}
M.~Frid-Adar, I.~Diamant, E.~Klang, M.~Amitai, J.~Goldberger, and H.~Greenspan,
  ``Gan-based synthetic medical image augmentation for increased cnn
  performance in liver lesion classification,'' \emph{Neurocomputing}, vol.
  321, pp. 321--331, 2018.

\bibitem{chen2020residual}
X.~Chen, L.~Yao, and Y.~Zhang, ``Residual attention u-net for automated
  multi-class segmentation of covid-19 chest ct images,'' \emph{arXiv preprint
  arXiv:2004.05645}, 2020.

\bibitem{chen2020momentum}
X.~Chen, L.~Yao, T.~Zhou, J.~Dong, and Y.~Zhang, ``Momentum contrastive
  learning for few-shot covid-19 diagnosis from chest ct images,'' \emph{arXiv
  preprint arXiv:2006.13276}, 2020.

\bibitem{zhu2017unpaired}
J.-Y. Zhu, T.~Park, P.~Isola, and A.~A. Efros, ``Unpaired image-to-image
  translation using cycle-consistent adversarial networks,'' in
  \emph{Proceedings of the IEEE international conference on computer vision},
  2017, pp. 2223--2232.

\bibitem{liu2017unsupervised}
M.-Y. Liu, T.~Breuel, and J.~Kautz, ``Unsupervised image-to-image translation
  networks,'' \emph{Advances in neural information processing systems},
  vol.~30, pp. 700--708, 2017.

\bibitem{goodfellow2014generative}
I.~Goodfellow, J.~Pouget-Abadie, M.~Mirza, B.~Xu, D.~Warde-Farley, S.~Ozair,
  A.~Courville, and Y.~Bengio, ``Generative adversarial nets,'' in
  \emph{Advances in neural information processing systems}, 2014, pp.
  2672--2680.

\bibitem{gulrajani2017improved}
I.~Gulrajani, F.~Ahmed, M.~Arjovsky, V.~Dumoulin, and A.~C. Courville,
  ``Improved training of wasserstein gans,'' in \emph{Advances in neural
  information processing systems}, 2017, pp. 5767--5777.

\bibitem{yang2018low}
Q.~Yang, P.~Yan, Y.~Zhang, H.~Yu, Y.~Shi, X.~Mou, M.~K. Kalra, Y.~Zhang,
  L.~Sun, and G.~Wang, ``Low-dose ct image denoising using a generative
  adversarial network with wasserstein distance and perceptual loss,''
  \emph{IEEE transactions on medical imaging}, vol.~37, no.~6, pp. 1348--1357,
  2018.

\bibitem{isola2017image}
P.~Isola, J.-Y. Zhu, T.~Zhou, and A.~A. Efros, ``Image-to-image translation
  with conditional adversarial networks,'' in \emph{Proceedings of the IEEE
  conference on computer vision and pattern recognition}, 2017, pp. 1125--1134.

\bibitem{mirza2014conditional}
M.~Mirza and S.~Osindero, ``Conditional generative adversarial nets,''
  \emph{arXiv preprint arXiv:1411.1784}, 2014.

\bibitem{li2019storygan}
Y.~Li, Z.~Gan, Y.~Shen, J.~Liu, Y.~Cheng, Y.~Wu, L.~Carin, D.~Carlson, and
  J.~Gao, ``Storygan: A sequential conditional gan for story visualization,''
  in \emph{Proceedings of the IEEE Conference on Computer Vision and Pattern
  Recognition}, 2019, pp. 6329--6338.

\bibitem{ledig2017photo}
C.~Ledig, L.~Theis, F.~Husz{\'a}r, J.~Caballero, A.~Cunningham, A.~Acosta,
  A.~Aitken, A.~Tejani, J.~Totz, Z.~Wang \emph{et~al.}, ``Photo-realistic
  single image super-resolution using a generative adversarial network,'' in
  \emph{Proceedings of the IEEE conference on computer vision and pattern
  recognition}, 2017, pp. 4681--4690.

\bibitem{zhang2019noise}
T.~Zhang, J.~Cheng, H.~Fu, Z.~Gu, Y.~Xiao, K.~Zhou, S.~Gao, R.~Zheng, and
  J.~Liu, ``Noise adaptation generative adversarial network for medical image
  analysis,'' \emph{IEEE Transactions on Medical Imaging}, vol.~39, no.~4, pp.
  1149--1159, 2019.

\bibitem{nie2017medical}
D.~Nie, R.~Trullo, J.~Lian, C.~Petitjean, S.~Ruan, Q.~Wang, and D.~Shen,
  ``Medical image synthesis with context-aware generative adversarial
  networks,'' in \emph{International Conference on Medical Image Computing and
  Computer-Assisted Intervention}.\hskip 1em plus 0.5em minus 0.4em\relax
  Springer, 2017, pp. 417--425.

\bibitem{zhang2019skrgan}
T.~Zhang, H.~Fu, Y.~Zhao, J.~Cheng, M.~Guo, Z.~Gu, B.~Yang, Y.~Xiao, S.~Gao,
  and J.~Liu, ``Skrgan: Sketching-rendering unconditional generative
  adversarial networks for medical image synthesis,'' in \emph{International
  Conference on Medical Image Computing and Computer-Assisted
  Intervention}.\hskip 1em plus 0.5em minus 0.4em\relax Springer, 2019, pp.
  777--785.

\bibitem{xue2018segan}
Y.~Xue, T.~Xu, H.~Zhang, L.~R. Long, and X.~Huang, ``Segan: Adversarial network
  with multi-scale l 1 loss for medical image segmentation,''
  \emph{Neuroinformatics}, vol.~16, no. 3-4, pp. 383--392, 2018.

\bibitem{dong2019neural}
N.~Dong, M.~Xu, X.~Liang, Y.~Jiang, W.~Dai, and E.~Xing, ``Neural architecture
  search for adversarial medical image segmentation,'' in \emph{International
  Conference on Medical Image Computing and Computer-Assisted
  Intervention}.\hskip 1em plus 0.5em minus 0.4em\relax Springer, 2019, pp.
  828--836.

\bibitem{khosravan2019pan}
N.~Khosravan, A.~Mortazi, M.~Wallace, and U.~Bagci, ``Pan: Projective
  adversarial network for medical image segmentation,'' in \emph{International
  Conference on Medical Image Computing and Computer-Assisted
  Intervention}.\hskip 1em plus 0.5em minus 0.4em\relax Springer, 2019, pp.
  68--76.

\bibitem{arjovsky2017wasserstein}
M.~Arjovsky, S.~Chintala, and L.~Bottou, ``Wasserstein gan,'' \emph{arXiv
  preprint arXiv:1701.07875}, 2017.

\bibitem{zhou2018unet++}
Z.~Zhou, M.~M.~R. Siddiquee, N.~Tajbakhsh, and J.~Liang, ``Unet++: A nested
  u-net architecture for medical image segmentation,'' in \emph{Deep Learning
  in Medical Image Analysis and Multimodal Learning for Clinical Decision
  Support}.\hskip 1em plus 0.5em minus 0.4em\relax Springer, 2018, pp. 3--11.

\bibitem{ronneberger2015u}
O.~Ronneberger, P.~Fischer, and T.~Brox, ``U-net: Convolutional networks for
  biomedical image segmentation,'' in \emph{International Conference on Medical
  image computing and computer-assisted intervention}.\hskip 1em plus 0.5em
  minus 0.4em\relax Springer, 2015, pp. 234--241.

\bibitem{long2015fully}
J.~Long, E.~Shelhamer, and T.~Darrell, ``Fully convolutional networks for
  semantic segmentation,'' in \emph{Proceedings of the IEEE conference on
  computer vision and pattern recognition}, 2015, pp. 3431--3440.

\bibitem{he2016deep}
K.~He, X.~Zhang, S.~Ren, and J.~Sun, ``Deep residual learning for image
  recognition,'' in \emph{Proceedings of the IEEE CVPR}, 2016, pp. 770--778.

\bibitem{ioffe2017batch}
S.~Ioffe, ``Batch renormalization: Towards reducing minibatch dependence in
  batch-normalized models,'' in \emph{Advances in neural information processing
  systems}, 2017, pp. 1945--1953.

\bibitem{huang2017densely}
G.~Huang, Z.~Liu, L.~Van Der~Maaten, and K.~Q. Weinberger, ``Densely connected
  convolutional networks,'' in \emph{Proceedings of the IEEE conference on
  computer vision and pattern recognition}, 2017, pp. 4700--4708.

\bibitem{kather2019predicting}
J.~N. Kather, J.~Krisam, P.~Charoentong, T.~Luedde, E.~Herpel, C.-A. Weis,
  T.~Gaiser, A.~Marx, N.~A. Valous, D.~Ferber \emph{et~al.}, ``Predicting
  survival from colorectal cancer histology slides using deep learning: A
  retrospective multicenter study,'' \emph{PLoS medicine}, vol.~16, no.~1, p.
  e1002730, 2019.

\bibitem{wang2017chestx}
X.~Wang, Y.~Peng, L.~Lu, Z.~Lu, M.~Bagheri, and R.~M. Summers, ``Chestx-ray8:
  Hospital-scale chest x-ray database and benchmarks on weakly-supervised
  classification and localization of common thorax diseases,'' in
  \emph{Proceedings of the IEEE conference on computer vision and pattern
  recognition}, 2017, pp. 2097--2106.

\bibitem{tschandl2018ham10000}
P.~Tschandl, C.~Rosendahl, and H.~Kittler, ``The ham10000 dataset, a large
  collection of multi-source dermatoscopic images of common pigmented skin
  lesions,'' \emph{Scientific data}, vol.~5, p. 180161, 2018.

\bibitem{kermany2018identifying}
D.~S. Kermany, M.~Goldbaum, W.~Cai, C.~C. Valentim, H.~Liang, S.~L. Baxter,
  A.~McKeown, G.~Yang, X.~Wu, F.~Yan \emph{et~al.}, ``Identifying medical
  diagnoses and treatable diseases by image-based deep learning,'' \emph{Cell},
  vol. 172, no.~5, pp. 1122--1131, 2018.

\bibitem{setio2017validation}
A.~A.~A. Setio, A.~Traverso, T.~De~Bel, M.~S. Berens, C.~van~den Bogaard,
  P.~Cerello, H.~Chen, Q.~Dou, M.~E. Fantacci, B.~Geurts \emph{et~al.},
  ``Validation, comparison, and combination of algorithms for automatic
  detection of pulmonary nodules in computed tomography images: the luna16
  challenge,'' \emph{Medical image analysis}, vol.~42, pp. 1--13, 2017.

\bibitem{yap2017automated}
M.~H. Yap, G.~Pons, J.~Mart{\'\i}, S.~Ganau, M.~Sent{\'\i}s, R.~Zwiggelaar,
  A.~K. Davison, and R.~Mart{\'\i}, ``Automated breast ultrasound lesions
  detection using convolutional neural networks,'' \emph{IEEE journal of
  biomedical and health informatics}, vol.~22, no.~4, pp. 1218--1226, 2017.

\bibitem{bilic2019liver}
P.~Bilic, P.~F. Christ, E.~Vorontsov, G.~Chlebus, H.~Chen, Q.~Dou, C.-W. Fu,
  X.~Han, P.-A. Heng, J.~Hesser \emph{et~al.}, ``The liver tumor segmentation
  benchmark (lits),'' \emph{arXiv preprint arXiv:1901.04056}, 2019.

\bibitem{xu2019efficient}
X.~Xu, F.~Zhou, B.~Liu, D.~Fu, and X.~Bai, ``Efficient multiple organ
  localization in ct image using 3d region proposal network,'' \emph{IEEE
  transactions on medical imaging}, vol.~38, no.~8, pp. 1885--1898, 2019.

\bibitem{salimans2016improved}
T.~Salimans, I.~Goodfellow, W.~Zaremba, V.~Cheung, A.~Radford, and X.~Chen,
  ``Improved techniques for training gans,'' \emph{Advances in neural
  information processing systems}, vol.~29, pp. 2234--2242, 2016.

\bibitem{heusel2017gans}
M.~Heusel, H.~Ramsauer, T.~Unterthiner, B.~Nessler, and S.~Hochreiter, ``Gans
  trained by a two time-scale update rule converge to a local nash
  equilibrium,'' in \emph{Advances in neural information processing systems},
  2017, pp. 6626--6637.

\bibitem{wang2016generative}
X.~Wang and A.~Gupta, ``Generative image modeling using style and structure
  adversarial networks,'' in \emph{European conference on computer
  vision}.\hskip 1em plus 0.5em minus 0.4em\relax Springer, 2016, pp. 318--335.

\bibitem{kingma2014adam}
D.~P. Kingma and J.~Ba, ``Adam: A method for stochastic optimization,''
  \emph{arXiv preprint arXiv:1412.6980}, 2014.

\bibitem{zhang2019adversarial}
X.~Zhang, L.~Yao, and F.~Yuan, ``Adversarial variational embedding for robust
  semi-supervised learning,'' in \emph{Proceedings of the 25th ACM SIGKDD
  International Conference on Knowledge Discovery \& Data Mining}, 2019, pp.
  139--147.

\bibitem{yao2016learning}
L.~Yao, F.~Nie, Q.~Z. Sheng, T.~Gu, X.~Li, and S.~Wang, ``Learning from less
  for better: semi-supervised activity recognition via shared structure
  discovery,'' in \emph{Proceedings of the 2016 ACM International Joint
  Conference on Pervasive and Ubiquitous Computing}, 2016, pp. 13--24.

\end{thebibliography}

\end{document}